\documentclass[reprint]{revtex4-2}
\usepackage[american]{babel}
\usepackage[utf8]{inputenc}
\usepackage{amsmath}
\usepackage{amssymb}
\usepackage{graphicx}
\usepackage{hyperref}
\usepackage{xcolor}
\usepackage[normalem]{ulem}
\graphicspath{{Figuras_Antigas/}}
 \newcommand{\be}{\begin{equation}}
\newcommand{\ee}{\end{equation}}
\begin{document}

\title{An entropy-based, scale-dependent Centrality}

\author{Lucas da Rocha Schwengber}
\email{lucas.schwengber@ufrgs.br, lucaschwengber@gmail.com}%, lucas.schwengber@impa.br}
\affiliation{Instituto de Matem\'atica e Estatística\\
Universidade Federal do Rio Grande do Sul\\
91501-970, Porto Alegre, Brazil}
 
\author{Sandra D. Prado and Silvio R. Dahmen}%
 \email{sandra.prado@ufrgs.br, silvio.dahmen@ufrgs.br}
\affiliation{Instituto de F\'{\i}sica\\
Universidade Federal do Rio Grande do Sul\\
91501-970, Porto Alegre, Brazil}
%
%\author{Silvio R. Dahmen}%
% \email{silvio.dahmen@ufrgs.br}
%\affiliation{Instituto de F\'{\i}sica, UFRGS}
%

\date{\today}
\begin{abstract}
In this article we introduce an entropy-based, scale-dependent centrality that is evaluated as the Shannon entropy of the distribution at time $t$ of a continuous-time random walk on a network. It ranks nodes as a function of $t$, which acts as a parameter and defines the scale of the network. It is able to capture well-known centralities such as degree, eigenvector and closeness depending on the range of $t$. We compare it with the broad class of total $f$-communicability centralities, of which both Katz centrality and total communicability are particular cases.
\end{abstract}

\maketitle

\section{Introduction}
For a variety of applications, the evaluation of the relative importance of nodes in a network is a topic of great interest. Since one expects networks may behave differently at different length scales, as the size and complexity of network data grow one
needs to refine standard centrality measures such as degree, betweenness, eigenvector centrality and closeness. New approaches that take into account the different behaviours at a local and a global level thus become necessary.

As discussed in \cite{BORGATTI200555} different centrality measures make different assumptions about how information, or any other quantity of interest, flows in a network. One particular way to describe the propagation of information and thus determine the importance of a node in a network is via a random walk. This has been traditionally used to define
several centrality measures, as for example the well-known PageRank algorithm \cite{ilprints422, Delvenne_2011}. More recent approaches employ the Shannon entropy of a constrained random walk as a way of measuring the uncertainty of the position of a signal traveling along the network \cite{TUTZAUER2007249,NIKOLAEV2015154,split&transfer}.

Another venue explores the idea of a  diffusion-based dynamics that is continuous in time but discrete in space. The idea is  to define a centrality that maps a node position relative to its neighborhood in terms of its diffusion up to time $t$ \cite{PhysRevResearch.2.033104}. In this scenario $t$ plays the role of a scaling parameter that controls how local the centrality is. Small values of $t$ are associated with a local conception of centrality, whereas large values with a global one.

The use of $t$ of a diffusion process as a scale parameter has provided remarkable results when it comes to recovering information about network data across different scales. 
In particular, it was used in \cite{DiffusionMaps} to capture a multi-scale idea of the geometry of a data-set. In the context of community detection, it was shown that by using both discrete and continuous time Markov chains on a graph, one is able to measure the stability of a given partition of the graph using $t$ as a scale parameter~\cite{Stability_Com, zoom_lens}. This provides a unified criteria to find optimal partitions over different scales.

In this paper we define an entropy-based, scale-dependent centrality $C^H(t)$ that unifies these approaches, that is a centrality based on the entropy of a random walk that is continuous in time. The centrality has two interpretations: it measures the uncertainty associated with the position of a random walker at a given $t$ as well as the time necessary for the convergence of the diffusion to its limiting distribution. We evaluate these by using the Shannon entropy of the distribution at diffusion time $t$.

%The parameter $t$ has a physical interpretation: it measures the uncertainty associated with the position of a random walker at a given $t$ as well as the time necessary for the convergence of the diffusion to its limiting distribution. We evaluate these by using the Shannon entropy of the distribution at diffusion time $t$. 

$C^H(t)$ can be interpreted in the context of social networks as well as in a purely physical context. For small values of $t$ it yields a ranking similar to Katz and other path-counting centralities, interpolating between degree and eigenvector centrality. For intermediate and larger values of $t$ it exhibits a significantly different behavior by going beyond eigenvector centrality and becoming related to closeness. Moreover, unlike \cite{PhysRevResearch.2.033104}, it is both continuous and differentiable with respect to $t$, making it easier to obtain rigorous results. In particular, we show that as $t \to 0$, $C^H(t)$ yields the same node rank as degree.

This article is organized as follows: in section II we define $C^H(t)$ in terms of Shannon entropy and discuss its interpretations. In section III we apply this centrality to a toy model and to some known networks of interest, and show its relation to other standard concepts of centrality. In section IV we establish some theoretical parallels with similar centralities, and give a theoretical result which establishes that for small $t$, $C^H(t)$ is equivalent to degree centrality. In section V we outline our conclusions and future perspectives.

\section{Entropy-based Centrality}

\subsection{Diffusion on Graphs}
Let $G = (V,E)$ be a connected and undirected graph with $N$ nodes. Consider a diffusion process on $G$ with total mass equal to unity described by: %\lucas{Fonte para a difusão}
\begin{equation}
 \frac{d \mathbf{p}(t)}{dt}  = -L~\mathbf{p}(t) \qquad 
 \mbox{with}\qquad
 \mathbf{p}(0)  = \mathbf{p}_0\,.
 \label{diffeq}
\end{equation}
Here the vector $\mathbf{p}(t)= [p_1(t) \dots  p_{N}(t)]^T$ encodes the density of some quantity at each node $i=1,2,\cdots, N$ of $G$ at time $t$. $L$ is the Laplacian matrix defined through $L = D - A$, where $A$ is the adjacency matrix and $D$ is the diagonal matrix whose entries are $d_{ii} = \mathrm{degree}(i)$. Equation \eqref{diffeq} constitutes a linear system of ordinary differential equations whose solutions are of the form:
\begin{equation}
\mathbf{p}(t) = e^{-tL} ~ \mathbf{p}_0\,.
\label{diffsol}
\end{equation}
We restricted ourselves to the case where $\mathbf{p}_0 = \delta_{ij}$, that is we start out with a unit mass at node $i$, the starting point of the process. In this case we denote the solution of (\ref{diffeq}) with this initial condition as $\mathbf{p}(t|i) = [p_1(t|i) \dots  p_{N}(t|i)]^T$ . 

\begin{figure}
    \includegraphics[scale=0.4]{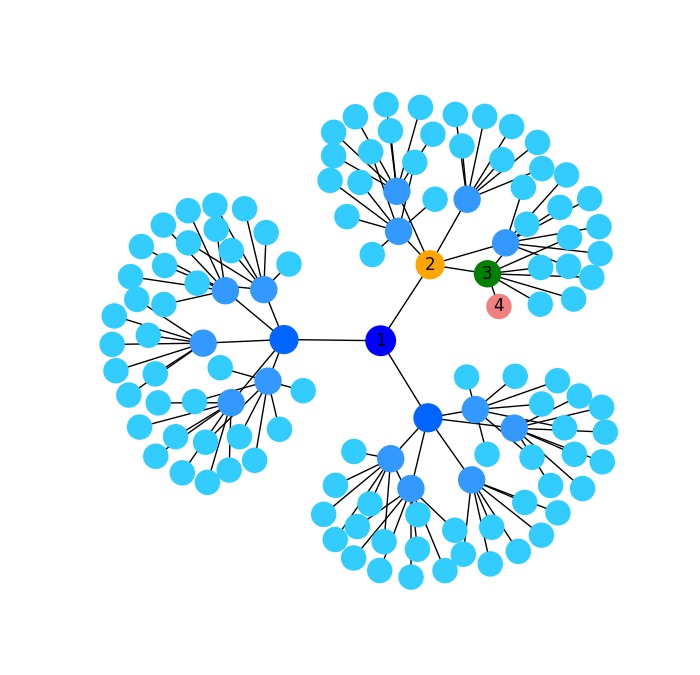}
    \caption{A tree-like network in which the root node $v_1$ (dark blue dot) generates 3 sprouts,
    one of which is $v_2$ (yellow dot). Each node of the first generation gives rise to $5$ new nodes, one of which is $v_3$ (green dot). Each node of this generation generates $7$ nodes nodes ({\it e.g.}, $v_4$, pink dot). The number of nodes is $N=124$.}
    \label{fig:tree}
\end{figure}

As $t$ becomes asymptotically large the solutions of \eqref{diffsol} converge to a uniform equilibrium, $p^e = [1/N \dots 1/N]^T$, with a common exponential rate regardless of the initial point $i$, although different starting points will make this convergence faster or slower for small and intermediate values of $t$. To see this consider as an example the graph represented by the tree in figure (\ref{fig:tree}). A process starting at the central node $v_1$ (dark blue dot) will initially evolve toward equilibrium much faster than one starting at node $v_4$ (pink dot) since $v_1$ is more connected with the rest of the network. However, as will be shown in section III, determining the differences as to how fast diffusion processes starting at $v_1,v_2, v_3$ and $v_4$ converge to equilibrium is more subtle. These differences between the speed of convergence for small and intermediate time scales is one of the main motivations behind the definition of the centrality presented below.

\subsection{Centrality}

The Entropy-Centrality $C^H(t)$ of node $i$ at time $t$ is defined as:
\begin{equation}
\begin{split}
    C^H_i(t) &= -\frac{1}{\log_2(N)\,}\sum_{j=1}^N p_j(t|i) \log_2(p_j(t|i)) .
\end{split}
\label{CHdef}
\end{equation}
This definition is simply the $t$-dependent Shannon formula for the normalized entropy of the distribution of a process that starts at node $i$ (see \cite{THOMASCOVER} for a discussion of Shannon entropy).%, and quantifies the progress toward equilibrium up to time $t$. 

Given that the Shannon entropy of a probability distribution over $N$ possible states reaches its maximum value $\log_2(N)$ bits for a uniform distribution $p^e$, we have:
\begin{equation}
\lim_{t \to \infty}C^H_i(t) = 1, \quad \forall i \in G.
\end{equation}
On the other hand, the distribution of a point mass at a single point has an entropy equal to $0$ bits:
\begin{equation}
C^H_i(0)=0, \quad \forall i \in G.
\end{equation}
For any given finite time $t=[0,\infty)$,~ equation (\ref{CHdef}) has a value between $0$ and $1$ and roughly measures how much the process starting at node $i$ has evolved up to time $t$.

This quantity can be interpreted in two different ways: on the one hand it can be seen as a state function that shows, for a given $t$, how close a diffusion process which started at node $i$ is to its equilibrium distribution. On the other hand, if we want to detect a particle whose random movement on the graph is governed by the transition rates $-L_{ij}$, the uncertainty of its position at time $t$ will depend on time elapsed since it started. This quantity can thus be interpreted as the uncertainty of the particle position at time $t$ given that it was  at position $i$ at $t=0$. % \lucas{(see \cite{THOMASCOVER} for an information-theoretical discussion of the Shannon Entropy)}.
For small $t>0$, the particle will have a small probability of being detected far from its immediate neighborhood: in this case the larger the $\text{degree}(i)$ the larger the uncertainty of its position. 
As $t$ increases, broader neighborhoods become increasingly more relevant and the uncertainty depends on the possible connections between $i$ and nodes farther away. 
The parameter $t$ thus plays a role similar to the dumping factor $\alpha$ of Katz centrality \cite{Katz1953} or the inverse temperature $\beta$ on the resolvent subgraph centrality \cite{matrix_functions}, subgraph centrality \cite{subgraph_cent} and total communicability \cite{total_comunicability}.

In the following sections we explore in more detail the properties of \eqref{CHdef}, its relationship to other well established centralities and how it performs when applied to known networks.

\section{Applications and Comparison with Other Centrality Measures}
\label{sec:applic}
We now present some applications of equation~(\ref{CHdef}), starting with a toy model and then discussing some known networks. We compare our results  with those of other commonly used centralities. 
We will see that despite the fact that the values $C^H(t)$ yields are related to degree, eigenvector centrality and closeness, % (a brief review of these centralities can be found in \cite{newman2010networks}),
with our method one is able to extract information about centrality using a unified scale. More specifically, our results suggest that this new centrality has a scale which is able to see farther than a class of walk-counting centrality measures, such as Katz centrality and total communicability (for the latter see section \ref{sec:applications}). In order to quantify their similarities we use the Spearman rank-order correlation. This quantity tells us how much a given centrality can be seen as a monotonic function of another centrality. 

Let $C_1,...,C_N$ and $C'_1,...,C'_N$ be the values at nodes $v_1,...,v_N$ of two different centrality measures $C$ and $C'$ on the same graph $G$. We set $R_{C_i}$ as the \textbf{rank} of the raw centrality score $C_i$ with respect the the other values of $C_j, j \neq i$. $R_{C'_i}$ is defined similarly. The Spearman rank correlation coefficient between $C$ and $C'$ (for a given $G$) is evaluated as:
\be
\label{eq:spearman}
\rho(C,C') = \frac{\text{cov}(\text{R}_C,\text{R}_{C'})}{\sigma_{\text{R}_C}\sigma_{\text{R}_{C'}}},
\ee
where $\text{cov}(\text{R}_C,\text{R}_{C'})$ is the covariance between $R_{C_1},...,R_{C_N}$ and $R_{C'_1},...,R_{C'_N}$ and $\sigma_{\text{R}_C}, \sigma_{\text{R}_C'}$ are the standard deviations associated with both respectively.

\label{parag1}
Henceforth we will indicate the supremum of the correlation between $C^H(t)$ and a given centrality measure $C'$ as \textit{correlation peaks} and their correspondent times as \textit{peak times}. More precisely, $\text{peak}_{C'} = \sup_{t > 0} \rho(C^H(t),C')$  %where $\rho(C^H(t),C')$ is the Spearman rank between $C^H(t)$ and some other centrality $C'$ 
for a graph  $G$.  The graph of $\rho(C^H(t),C')$ as a function of $t$ will be referred to as \textit{correlation curves}.

\subsection{A Toy Model}
\label{sec:toymod}
\begin{figure}
    \includegraphics[scale=0.4]{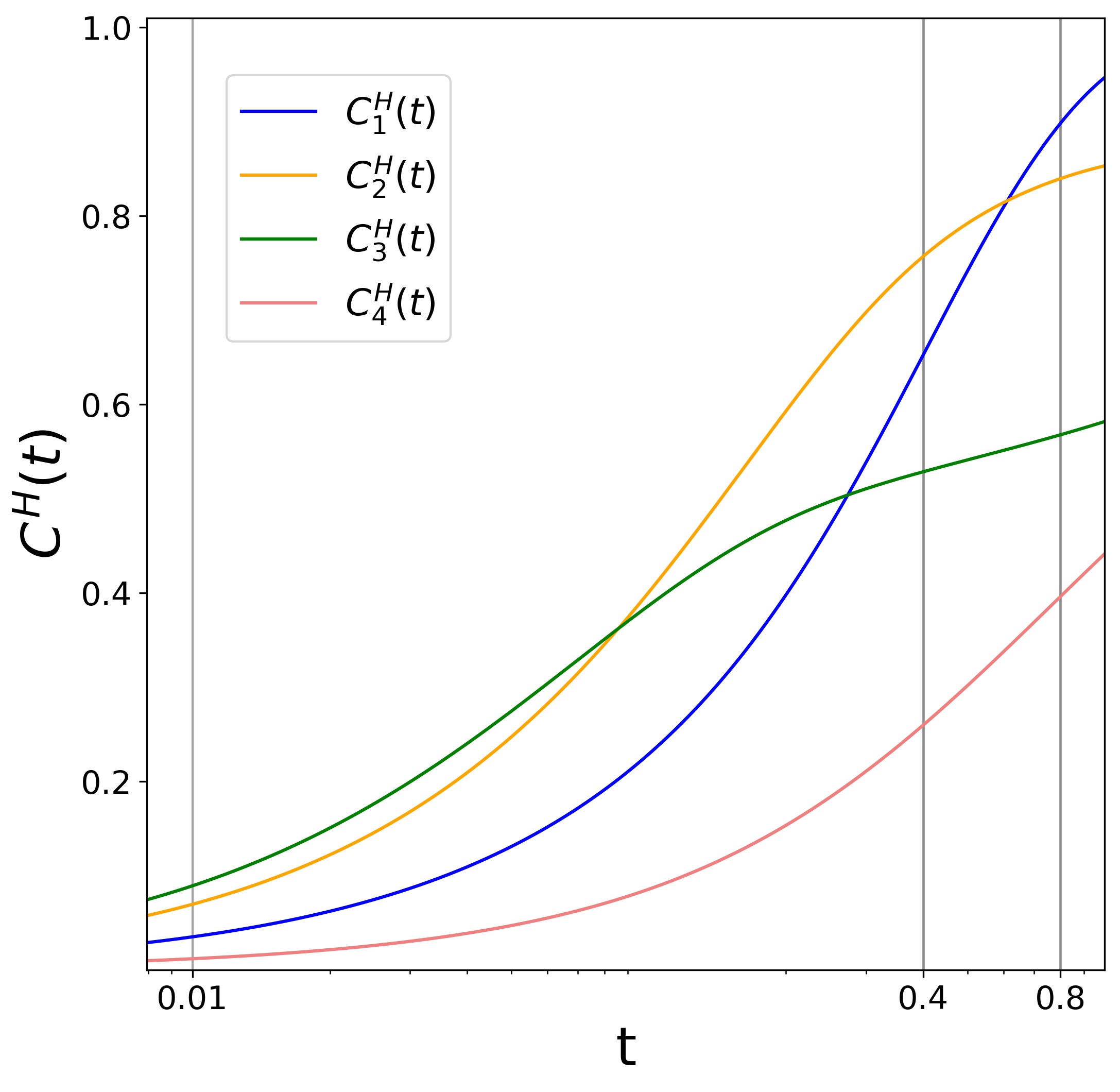}
    \caption{Evolution of $C^H_i(t)$ with respect to $t$ of the four nodes $v_1,v_2,v_3,v_4$ depicted in figure \ref{fig:tree}. The vertical gray lines represent the values of $t$ used in Table \ref{tab:t1}.} %Locally, $v_3$ is more central since it has the largest degree (8, against 6 and 3 for $v_2$ and $v_1$ respectively).}
    \label{fig:toymodelcurves}
\end{figure}

We first present a simple toy network in the form of a tree as depicted in Figure (\ref{fig:tree}). Table \ref{tab:t1} gives the values for degree, eigenvector and closeness centrality for the four types of node in this network: $v_1$ (dark blue), $v_2$ (yellow), $v_3$ (green) and $v_4$ (pink).
Each of the three centralities rank nodes differently, degree being the most local, closeness a more global measure and eigenvector centrality a measure for intermediate scales.
Figure (\ref{fig:toymodelcurves}) shows that for very small $t$ such as  $t=0.01$ indicated by a vertical line in this figure, $C^H_i(0.01)$ gives the same ranking as degree, but it changes with time. As an example,  $C^H_i(0.4)$  gives the same ranking as the eigenvector centrality, and for $C^H_i(0.8)$ we get the same ranking as closeness. The precise results are shown in Table \ref{tab:t1}. The
 Katz centrality has a scale parameter $\alpha$ which is able to reproduce degree, when $\alpha$ approaches, but is different from, $0$ and eigenvector centrality, when $\alpha$ approaches the inverse of the largest eigenvalue of $A$. However it fails to capture, on a larger scale, the fact $v_1$ is more central in the sense of closeness centrality. The same limitation applies to total communicability (for a definition see \eqref{tot_com} below). $C^H_i(t)$ on the other hand is able to reproduce three concepts of centrality depending on the scale defined uniquely by $t$ as it varies. 
{\small 
\begin{table}
\begin{tabular}{c|cccccc}
v & $C^H_i(0.01)$ & deg. & $C^H_i(0.4)$ & eigenv. & $C^H_i(0.8)$ & closeness \\ \hline
1     & 0.0352                           & 3                           & 0.6527                          & 0.2460                           & 0.8985                          & 0.3534                         \\
2     & 0.0697                           & 6                           & 0.7571                          & 0.3006                           & 0.8397                          & 0.3153                         \\
3     & 0.0894                           & 8                           & 0.5287                          & 0.1712                           & 0.5679                          & 0.2470                         \\
4     & 0.012                            & 1                           & 0.2599                          & 0.0467                           & 0.3963                          & 0.1984                         \\ %\bottomrule
\end{tabular}
\caption{Values of $C^H(t)$ at $t=0.01,\,0.4,\,0.8$ in comparison to degree, eigenvector centrality and closeness for nodes $v_1,v_2,v_3,v_4$ of Figure (\ref{fig:tree}). Note that up to a scaling factor, $C^H(t)$ reproduces approximately the values of degree and eigenvector for small and large $t$ respectively.}
\label{tab:t1}
\end{table}
}

\subsection{Applications to known Networks}
\label{sec:applications}
%In the following we illustrate our results by comparing  %(see figures \ref{fig:netcorr}, \ref{fig:netcorr_total} and \ref{fig:orbis_corr}, \ref{fig:orbis_corr_total} below), 
%$C^H_i(t)$ to the Katz Centrality and other walk-counting centralities such as total communicability.
%Moreover, there comes to a point where these centralities are not able to capture features beyond eigenvector centrality, while $C^H_i(t)$ does, as illustrated by the toy model in the preceding section.

In this section we apply $C^H$ to three known networks:  Zachary's Karate Club, the Netscience citation network of coauthorship compiled by M. Newman, and the  Stanford Geospatial Network Model (ORBIS) of the Roman World. We study the Spearman rank correlation with other centralities using 
$\rho(C,C')$ of equation (\ref{eq:spearman}) and present some graphs of $C^H_i(t)$ for  values of $t$ where there is a peak in correlation with degree, eigenvector and closeness.
The correlations curves are intended to shed light on the behavior of $C^H(t)$.

We compare $C^H(t)$ with the total communicability $C^T_i(\beta)$ of node $i$ defined as \cite{total_comunicability}
\be
\label{tot_com}
C^T_i(\beta) = \sum_{j=1}^n \sum_{k=0}^{\infty} \frac{\beta^k}{k!}(A^k)_{ij} = \sum_{j=1}^n \left(e^{\beta A}\right)_{ij}\,.
\ee
%That is, the sum of the $i$-th row of $\exp({\beta A})$.
The choice of total communicability instead of the widely known Katz centrality is due to the fact that it behaves more smoothly when the scale parameter approaches its limiting value.

\subsection*{Zachary Karate Club}
\begin{figure}
    \includegraphics[scale=0.4]{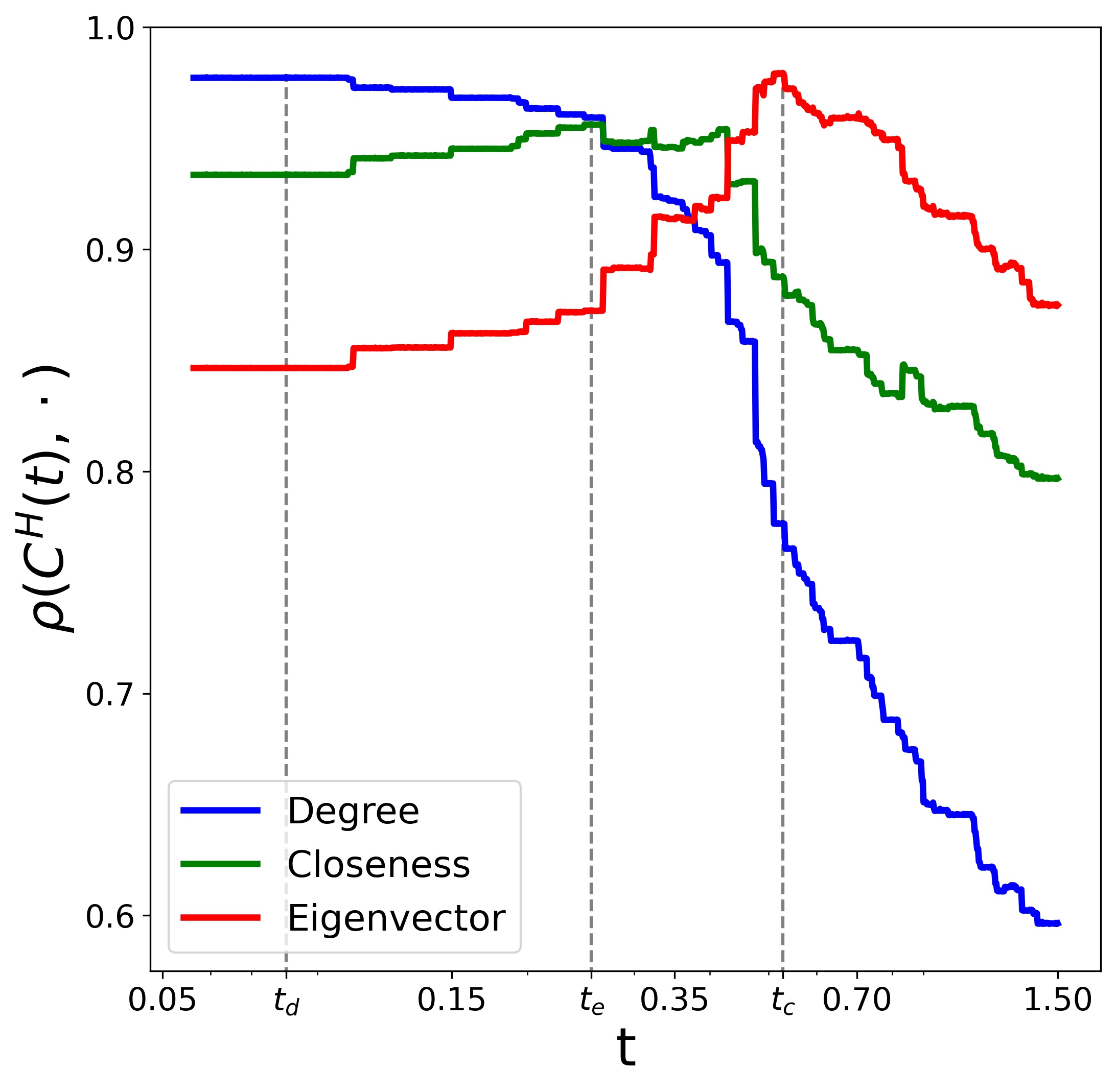}
    \caption{Spearman correlation coefficient between $C^H(t)$ and  degree, eigenvector and closeness. The vertical dashed lines indicate the values of $t$ for which there is a the peak in the correlation with degree ($t_d$), eigenvector centralidade ($t_e$) and closeness ($t_c$).}
    \label{fig:karatecorr}
\end{figure}

Our first application concerns the canonical Zachary Karate Club network. Figure (\ref{fig:karatecorr}) depicts the Spearman correlation between $C^H_i(t)$ and degree, closeness and eigenvector centrality. 
\begin{figure*}
    \includegraphics[scale=0.3]{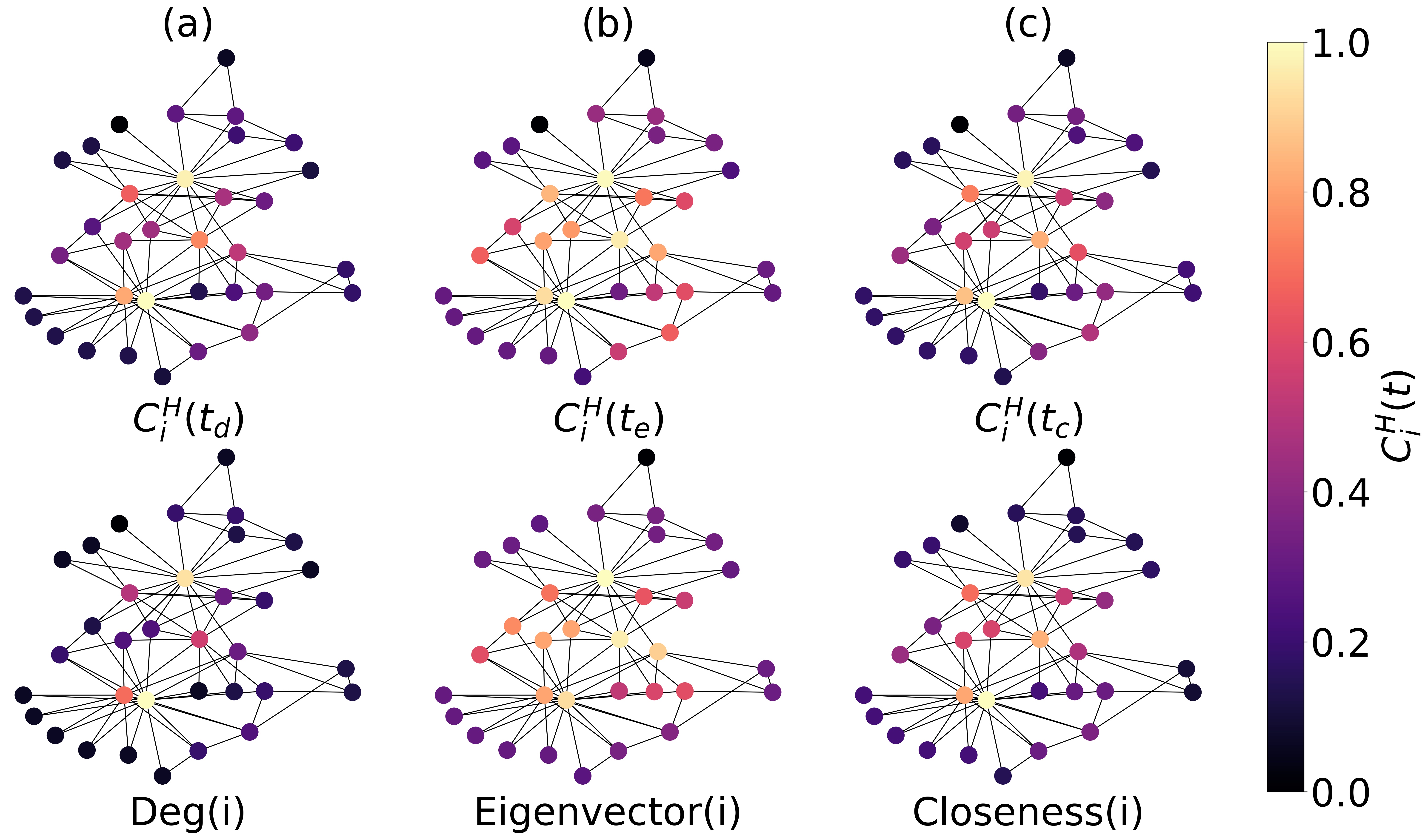}
    \caption{The upper section shows the relative values of $C^H(t)$ for $t_d$, $t_e$ and $t_c$. These are the times
    for which there is a peak in the correlation between $C^H(t)$ degree, eigenvector and closeness. The lower section shows the relative values of the corresponding centralities for the Karate Club network.}
    \label{fig:karatemap}
\end{figure*}
As it is shown in Figure  (\ref{fig:karatecorr}), the Spearman correlation with degree, closeness and eigenvector is overall very high with peaks above 0.9 for all them. This uniformly high correlation can be explained by the fact that these centralities are highly correlated on the Karate Club network. 
As for the peak times, their order seems to corroborate the idea that this is indeed a scale-dependent centrality. 
The peak with degree happens at a very small time parameter; 
the ones with closeness and eigenvector, which carry a more global measure, happen later. 

Figure (\ref{fig:karatemap}) shows that $C^H_i(t)$ reproduces node degree up to a constant. The peak values of $C^H_i(t)$ with respect to eigenvector and closeness are more interesting to analyze. The first thing to notice is that the relative values of $C^H_i(t)$ are generally more spread out which is a direct consequence of the fact that they all have the same limit. The peripheral nodes in $(4b)$ have a higher relative value in $C^H_i(t_e)$ than in eigenvector centrality. The same is valid for figure $(4c)$, but in this case they are slightly less spread out since $t_c$ occurs earlier. Note that in figure  (\ref{fig:karatemap}),  $t_d, t_e$ and $t_c$ denote peak times for degree, eigenvector and closeness centralities.

\subsection*{Netscience Coauthorship Network}
The second network we considered was Netscience, the graph of coauthorship of scientists working on network theory and experiment, as compiled by M. Newman in May 2006 from the bibliographies of \cite{Boccaletti2006} and \cite{Newman2003} with a few references added by hand.  The network is
weighted  as described in \cite{Newman2001}. For our purposes we considered only the giant component consisting of $N=379$ nodes.
\begin{figure}
    \includegraphics[scale=0.4]{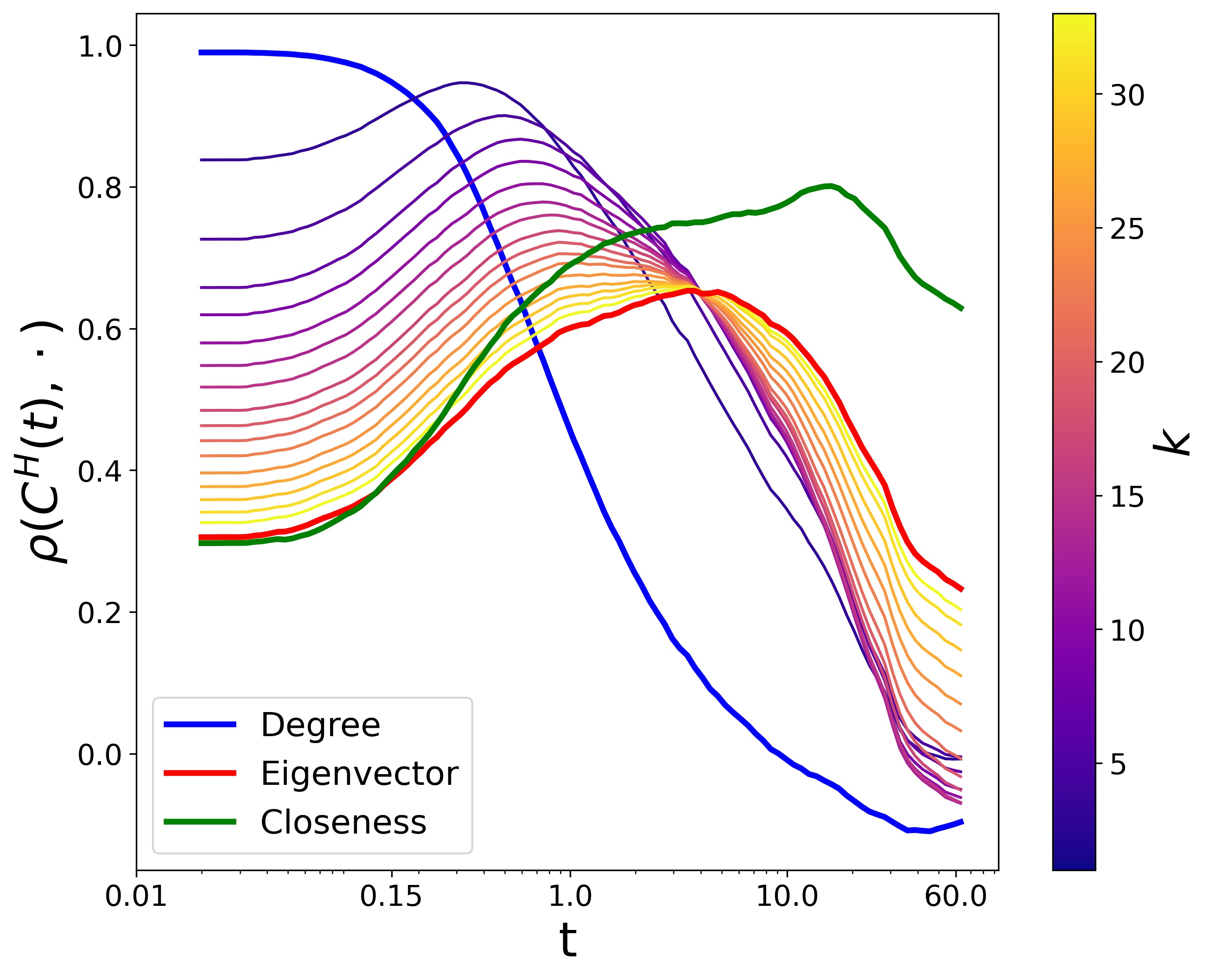}
    \caption{Spearman correlation coefficient between $C^H(t)$ and degree, eigenvector and closeness for the Netscience network. The purple-yellow scale on the right-hand side is the correlation of $C^H(t)$ with the sum of rows of the matrix powers $A^k$, The values of $k$, ranging from $k=1$ to $k=35$, are marked on this scale.}
    \label{fig:netcorr}
\end{figure}

\begin{figure}
    \includegraphics[scale=0.4]{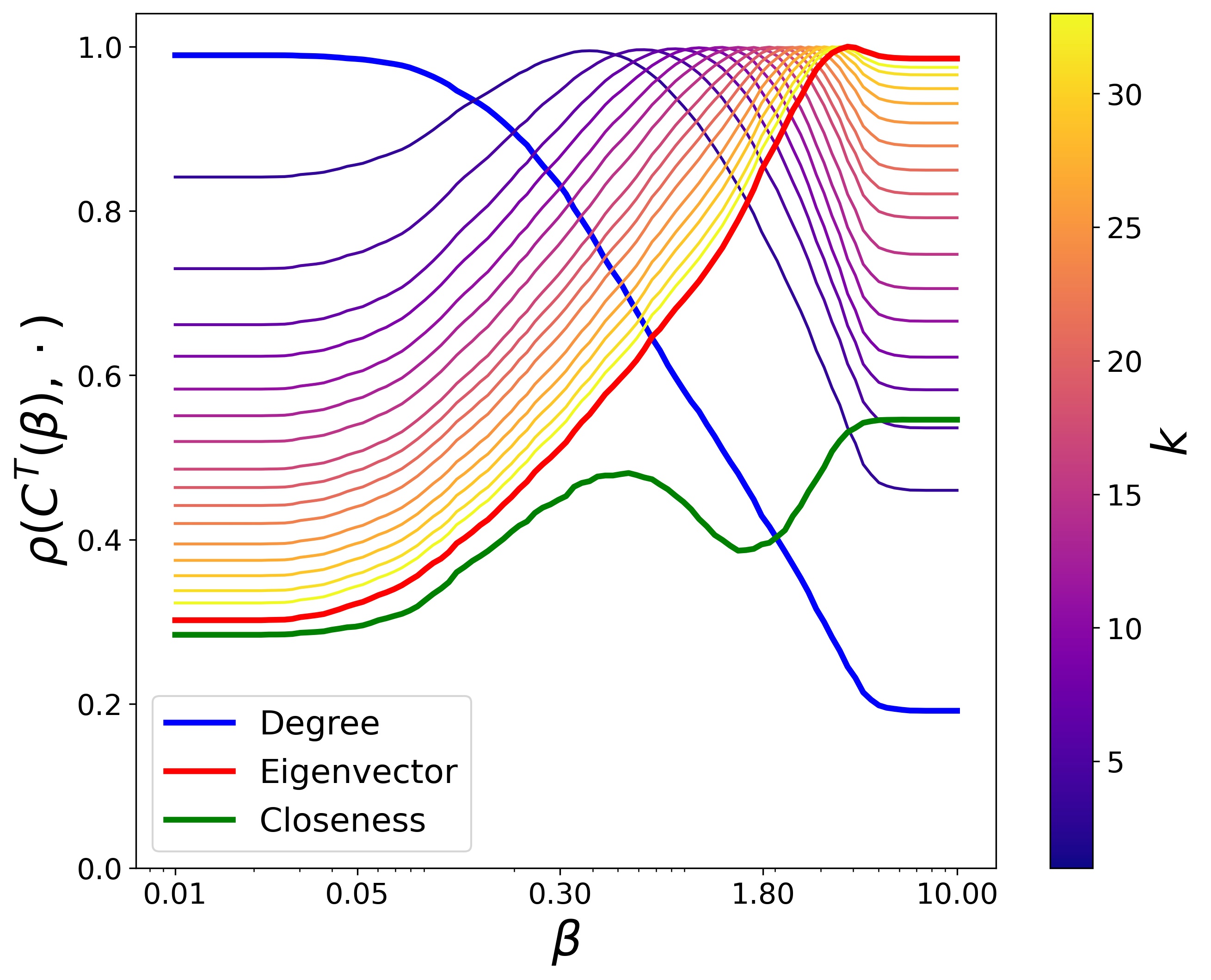}
    \caption{Spearman correlation coefficient between $C^T(\beta)$ and degree, eigenvector and closeness for the Netscience network. Compare to figure (\ref{fig:netcorr}).}.
    \label{fig:netcorr_total}
\end{figure}

As figure (\ref{fig:netcorr}) suggests, some key correlations such as closeness and eigenvector centrality are not as high as in the previous network. The order of the scales looks similar to the toy model, and they preserve the property that correlation with eigenvector centrality and closeness takes place after node degree, reinforcing the role of $t$ as a scale parameter. $C^H(t)$ starts with a very high correlation with node degree ($\rho = 0.989$), then presents a moderate correlation with eigenvector centrality ($\rho = 0.653$) and finally a high correlation with closeness ($\rho = 0.800$). To gain additional information, we plotted the correlation curves between $C^H(t)$ and the number of walks of length $k$ starting at a node $i$ (which is equal to the sum of the $i$-th of $A^k$). For $k=1$ this coincides with degree (blue curve in figures  \ref{fig:netcorr} and \ref{fig:netcorr_total}), whereas as $k \to \infty$ it yields the same rank as eigenvector centrality (red curve in Figures (\ref{fig:netcorr}) and (\ref{fig:netcorr_total})).  

The curves associated with intermediate values of $k$ show how the correlation changes from degree to eigenvector centrality. In figure (\ref{fig:netcorr_total}) the peak of the blue correlation curve happens as $t \to 0$ and the peak of the red correlation curve associated with eigenvector centrality is the limiting behavior of the scale parameter $\beta$.  As $\beta$ ranges from $0$ to $\infty$, $C^T(\beta)$ interpolates between degree (blue curve) and eigenvector centrality (red curve) \cite{matrix_limits}. In contrast, in figure \ref{fig:netcorr} as $t \to 0$ there is a high correlation with degree (blue curve) but the peak of correlation with eigenvector centrality (red curve) is not the limiting behavior of $C^H(t)$ as $t \to \infty$, but rather an intermediate step. After the peak of correlation with eigenvector there is also a substantial increase in correlation with closeness centrality (green curve) which does not happen with total communicability. This behavior is analogous to what was seen more sharply in the toy model.

This suggests that in some sense, $C^H$ extends the scale of Katz centrality and total communicability, which ends at eigenvector centrality, by correlating well with closeness afterwards. As Figure  (\ref{fig:netcorr_total}) shows, total communicability is unable to correlate well with closeness since its limiting behavior is restrited to eigenvector centrality. The cost of $C^H(t)$ ability to see further seems to be a not so high peak of correlation with eigenvector and the number of walks of length $k$, starting at a given node, as $k$ grows.

\subsection*{ORBIS: The Stanford Geospatial Network Model of the Roman World}
Next we consider a network formed by all existing land or see routes between cities of the Roman Empire around 200 CE. This network is generated by data compiled for the Stanford Geospatial Network Model of the Roman World (ORBIS). This network depicts $N=678$ major cities and in spite of being very complete (cost of travel, travel times, etc.) we worked only with the information on whether the locations were connected or not. 

This network is particularly suitable to the uncertainty of the random walk interpretation we gave to $C^H(t)$. The values of the centrality at each geographical location roughly indicate how hard it would be to capture someone who has disappeared at that given location and started moving randomly in the network up to time $t$. 

\begin{figure}
    \includegraphics[scale=0.4]{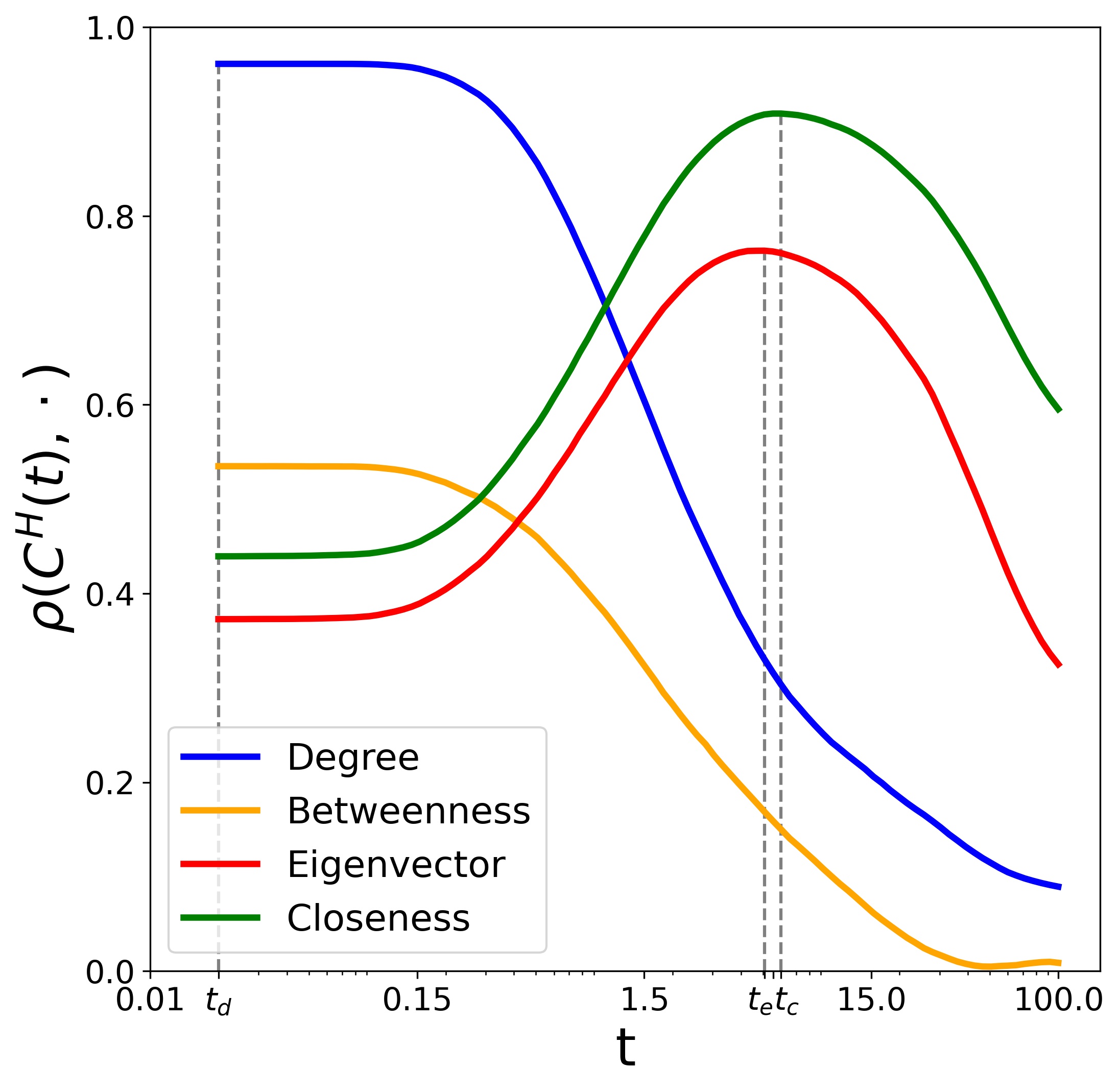}
    \caption{Spearman correlation coefficient between $C^H(t)$ and degree, betweenness, eigenvector and closeness for the ORBIS network. The vertical dashed lines indicate the time for which the peak in correlation with degree ($t_d$), eigenvector centrality ($t_e$) and closeness ($t_c$) takes place.}
    \label{fig:orbis_corr}
\end{figure}

Figure \ref{fig:orbis_corr} shows the Spearman rank correlation curves between $C^H(t)$ and degree, betweenness, eigenvector and closeness. We included betweenness (yellow curve) in this particular case to give a glimpse of why the correlation is not so meaningful for this centrality. Generally since $C^H_i(t)$ correlates almost perfectly with degree for small scales, it will inherit any correlation degree has with other centralities, and that seems to be the case with betweenness. The peak of correlation of $C^H_i(t)$ and betweenness occurs for very small values of $t$ as figure (\ref{fig:orbis_corr}) shows, and has  a similar magnitude ($\rho = 0.535$) as the correlation between degree and betweenness ($\rho = 0.572$). That together with the interpretation of $C^H(t)$, seems to indicate that $C^H(t)$ and betweenness are not deeply related beyond the typical correlation one might expect between centrality measures (see \cite{correl}).

An interesting feature of figures (\ref{fig:netcorr}) and (\ref{fig:orbis_corr}) is that in both of them the peak of correlation with closeness (indicated by the green line) happens after the one with eigenvector centrality, the same behavior we observed in the results for our toy model (see Table \ref{tab:t1}).  However as can be seen in figure (\ref{fig:karatecorr}) this may not be always the case.

\begin{figure}
    \includegraphics[scale=0.4]{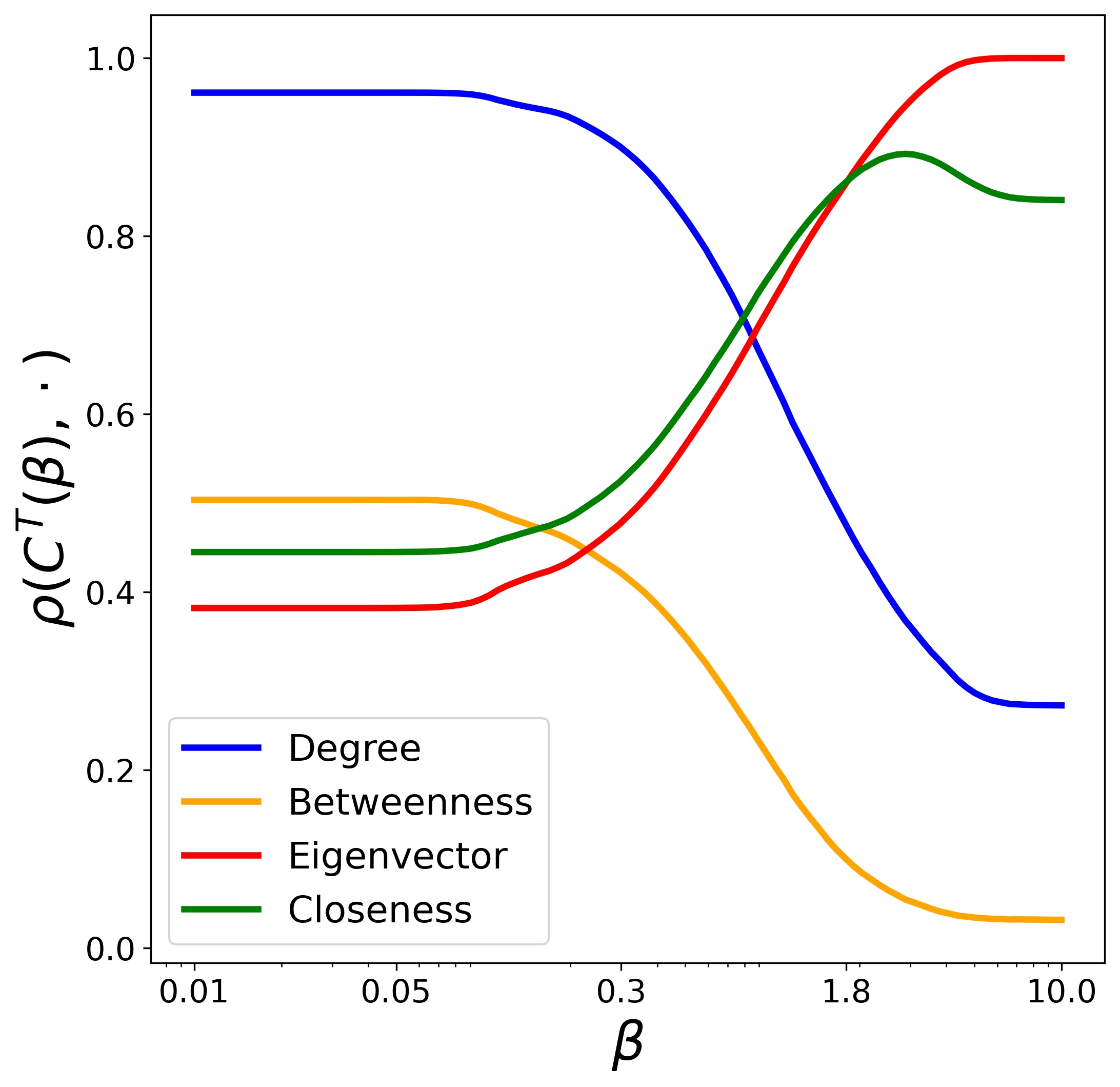}
    \caption{Spearman coefficient between $C^{T}(\beta)$  and degree, betweenness, eigenvector and closeness for the ORBIS network.}
    \label{fig:orbis_corr_total}
\end{figure}

Figures (\ref{fig:orbis_corr}) and (\ref{fig:orbis_corr_total}) are equivalent to figures (\ref{fig:netcorr}) and (\ref{fig:netcorr_total}) but for the ORBIS network. We did not include the correlation  with the number of steps of length $k$, shown in figures (\ref{fig:netcorr}) and (\ref{fig:netcorr_total}), for clarity. The contrast in behavior presented in figures (\ref{fig:netcorr}) and (\ref{fig:netcorr_total}) can also be seen in figures (\ref{fig:orbis_corr}) and (\ref{fig:orbis_corr_total}). $C^T(\beta)$ interpolates between degree and eigenvector as $\beta$ goes from $0$ to $\infty$ whereas $C^H(t)$ starts with a high correlation with degree, then a peak of correlation with eigenvector followed by a peak of correlation with closeness. Even though we have a high correlation between closeness and eigenvector centralities ($\rho = 0.841$) this correlation does not explain the peak in correlation that $C^H(t)$ exhibits with closeness ($\rho = 0.908$) as depicted in Figure \ref{fig:orbis_corr}.

\begin{figure*}
  \includegraphics[width=170mm]{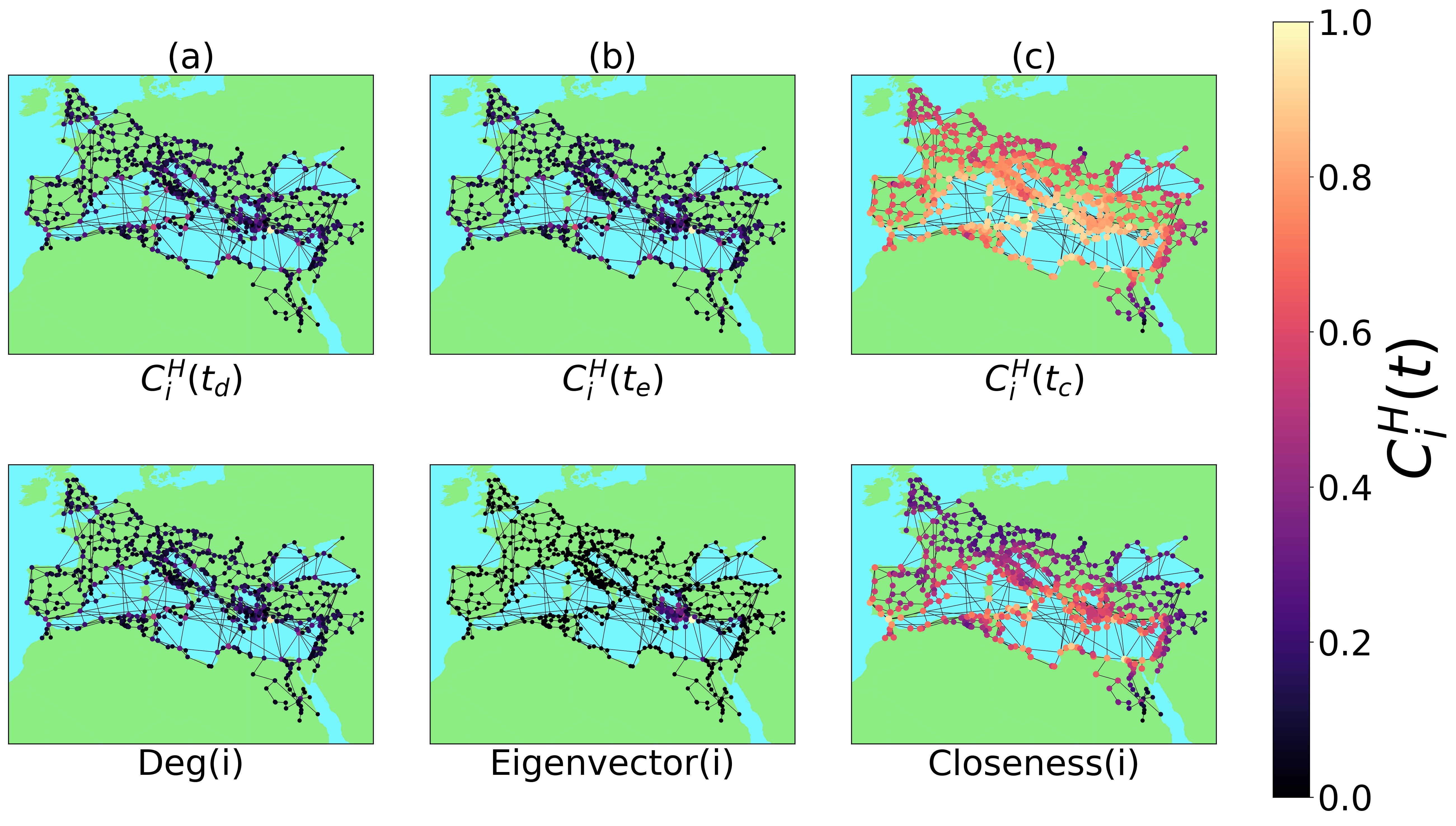}
  \caption{The upper section shows the relative values of the Entropy-Based Multiscale Centrality at different time values: $t_d$, $t_e$ and $t_c$ to time where the peaks of the correlation with degree, eigenvector and closeness centralities occur respectively. The lower section shows the relative values of the associated classical centralities for the Orbis network.}
  \label{fig:orbismap}
\end{figure*}

Figure (\ref{fig:orbismap}) depicts qualitative results for a much larger network corroborating the results for the Karate Club depicted in figure (\ref{fig:karatemap}). 
For small values of $t$, $C^H(t)$ is proportional to the node degree, as shown in figure (\ref{fig:orbismap}$a$). However, as it can be inferred from figure (\ref{fig:orbismap}$b$), whereas the eigenvector centrality is larger (in relative magnitude) for some localities around the Aegean Sea, the value $C^H(t_e)$ is more evenly spread throughout the European continent, where $t_e$ is the value that maximizes the correlation between $C^H(t)$ and the eigenvector centrality. 

\section{Relation with other centrality measures}

\subsection{Total $f$-Communicability}
$C^H_i(t)$ can be seen as some sort of nonlinear analog of the Katz centrality \cite{Katz1953} and total communicability \cite{total_comunicability} and more generally of total $f$-communicability centralities defined by equation (\ref{f_cent}) below \cite{matrix_limits,matrix_functions}. For instance, the $\alpha$-centrality \cite{alpha_cent} which differs from the Katz centrality only by an extra term that does not change the rank, is obtained by taking the sum of rows of the matrix defined by resolvent of $A$:
\be
\left(I-\alpha A\right)^{-1} = I + \alpha A + \alpha^2 A^2 + \cdots + \alpha^n A^n + \cdots
\ee
where $0 < \alpha < 1/\lambda_1$, $\lambda_1$ being the largest eigenvalue of $A$.
%The total communicability defined in (\ref{tot_com}) is given by the sum of the $i$th row of:
%\be
%e^{\beta A} = I + \beta A + \frac{(\beta A)^2}{2!} + \cdots + \frac{(\beta A)^n}{n!}  + \cdots
%\ee
%to be the centrality of node $i$.
%T
More generally as suggested by the general framework in \cite{matrix_functions} and further explored in \cite{matrix_limits}, one can take a series of the form

\be\label{matrix_f}
f(tA) = c_0 + c_1 tA + c_2 t^2 A^2 + ...\,,
\ee
with $c_i > 0,~ i=0,1,2,\cdots$ and where $f(x) = \sum_{k=0}^{\infty} c_k x^k$ has a positive (possibly infinite) radius of convergence $R_f$. If $R_f<\infty$ we additionally impose that $\lim_{t \to 1^-}f(t R_f) = +\infty$. Although \cite{matrix_functions} and \cite{matrix_limits} consider more than one way of using equation (\ref{matrix_f}) to derive centrality measures, we restrict ourselves to the case of the so-called total $f$-communicability:
\be\label{f_cent}
C^f(t)_i = \sum_{j=1}^n (f(tA))_{ij}
\ee
Which is just the sum of the $i$-th row of $f(tA)$. Both Katz centrality and total communicability are particular cases of equation(\ref{f_cent}). The former is obtained by using $f(x) = \sum_{k=0}^{+\infty} x^k$ whereas for the latter one uses $f(x) = \sum_{k=0}^{+\infty} x^k/k!$.

As shown in \cite[Theorem 5.1]{matrix_limits}, under the assumption that $A$ is primitive and $f$ is defined on the spectrum of $A$, as $t$ ranges from $0$ to its limiting value $t^* = R_f/\lambda_1$, the associated total $f$-communicability centrality interpolates between degree and eigenvector centrality. Thus the limiting behaviors in Figure (\ref{fig:netcorr_total}) and (\ref{fig:orbis_corr_total}) are expected not only for total communicability but also any total $f$-communicability under the given assumptions. 

$C^H(t)$ is also defined in terms of a matrix series expansion:
\be
e^{-tL} = I - t L + \frac{t^2 L^2}{2!}  + \cdots + \frac{(-t)^n L^n}{n!} + \cdots
\ee
However taking the row sums for this expansion would not make sense since under the assumption that $A$ is symmetric, all row sums are equal to $1$. Instead we use a non-linear approach by taking the entropy of each row, which then gives a meaningful way to distinguish nodes.

The applications above show that the behavior of $C^H_i(t)$ is similar to the total $f$-communicability centralities in that it resembles degree as $t \to 0$ and then, as $t$ grows, starts correlating better with eigenvector centrality. However differently from the total $f$-communicability centralities, the peak of correlation with eigenvector is not the limiting behavior of $C^H_i(t)$ but rather an intermediate step. It is as if $C^H(t)$ starts as a path counting centrality but changes its behavior to something more closeness-like as $t$ grows. The contrast between figures (\ref{fig:netcorr}) and (\ref{fig:netcorr_total}) and figures (\ref{fig:orbis_corr}) and (\ref{fig:orbis_corr_total}) showcases this distinction with total communicability representing the behavior of the total $f$-communicability. The cost of being able to reach larger scales seems to be that the correlation with eigenvector may not be very high as figure (\ref{fig:netcorr_total}) clearly shows. Below we show that the same result about the limiting behavior as $t \to 0$ of the total $f$-communicabilities hold for $C^H(t)$. However as the applications showed, the counterpart when $t \to \infty$ cannot be the same.

\subsection{An Explanation for the high correlation with degree}
\label{sec:exp_deg}

The results in the previous section were mostly qualitative, as we checked results by calculating correlations with particular graphs. In what follows we show that  in the case of degree at least, this result is rigorous.

\textbf{Proposition 1:} Given a connected and undirected Graph $G$ with $C^H(t)$ being its entropy-based scale-dependent centrality, there exists a $\delta>0$ such that, $\forall$  $t \in (0,\delta)$, $C^H(t)$ satisfies: $\mathrm{degree}(i)>\mathrm{degree}(j) \Rightarrow C^H_i(t) > C^H_j(t), \forall\, i,j\,\in\,G$. \\

This means that for sufficiently small $t$, $C^H_i(t)$ follows the ranking given by degree. The result can be improved to showing that actually for sufficiently small $t$, $C^H_i(t) \approx \kappa(t) \cdot \mathrm{degree}(i)$, where $\kappa(t)$ is a function of $t$ that does not depend on any particular node $i$. That explains why $C^H(t)$ is able to reproduce not only the rank of degree, but the relationship of its relative values among nodes. A proof of this proposition is given in the \hyperref[sec:appen]{Appendix}.

%\lucas{This result is analogous to the first part of Theorem 5.1 in \cite{matrix_limits}}.

\begin{figure}
    \includegraphics[scale=0.4]{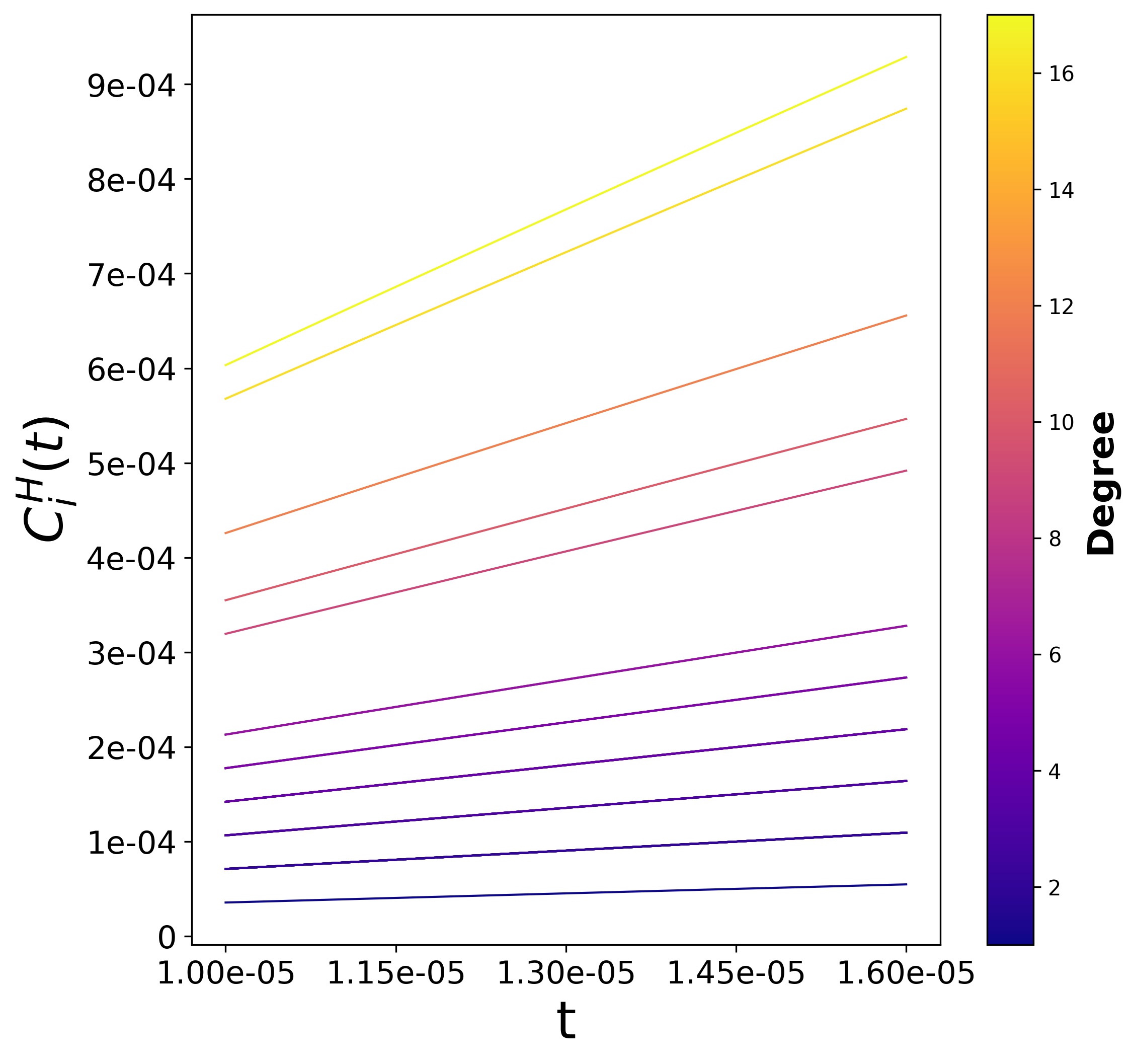}
    \caption{A small window of the entropy curves for nodes in the Karate Club network at very small time scales. Some curves associated with nodes with equal degree are overlapping.}
    \label{deg_corr}
\end{figure}

Figure \ref{deg_corr} illustrates the claim of Proposition $1$. The ranking of the $C^H_i(t)$ curves for a small window of $t$ is identical to the ranking of the degree of the respective nodes. The resemblance is so sharp that the curves for nodes which have the same degree are practically indistinguishable in the mentioned figure.

\section{Conclusion and further work}
In this work we showed how to derive a scale-dependent centrality %from the same ideas of diffusion as \cite{PhysRevResearch.2.033104} 
using the Shannon entropy of the diffusion process that starts at a given node. This centrality shares some similarities with standard centrality measures such as Katz-centrality \cite{alpha_cent,Katz1953} and more generally total $f$-communicability centralities \cite{matrix_limits, matrix_functions}, but has a remarkably different behavior both in the intermediate range of values of the scale parameter $t$, as well as when $t$ goes to its limiting value. In some way $C^H(t)$ is able to extend the scale  and capture a behavior which goes beyond eigenvector centrality. The cost paid is the decay in correlation with walks of length $k$ as $k \to \infty$. 
Our results were illustrated an interpreted with the help of a toy model and the known networks: Zachary's Karate Club, the Netscience Citation Network and the ORBIS map of the Roman World.

We remark that even though we have empirical evidence 
that $C^H_i(t)$ has a scale-dependent behavior that takes into account longer walks as $t$ grows, it is still not all clear what this relation is. From the results in section \ref{sec:applic} we conjecture it must be something fundamentally different from the way the total $f$-communicability centralities count walks but this is something still to be understood. \\

We thank W. Scheitel and V. Abraham from the ORBIS project for sharing their databank with us. We also would like to thank R. Misturini for the fruitful discussions on the Markov chain interpretation of $C^H_i(t)$ and Proposition 1.

\section{Appendix}
\label{sec:appen}

\textbf{Proposition 1:} Given a connected, undirected and weighted Graph $G$ with $C^H(t)$ being its entropy-based scale-dependent centrality, there exists a $\delta>0$ such that, $\forall$  $0<t<\delta$, $C^H(t)$ satisfies: $\mathrm{degree}(i)>\mathrm{degree}(j) \Rightarrow C^H_i(t) > C^H_j(t), \forall\, i,j\,\in\,G$. \\

\textbf{Proof of Proposition 1:}
Let $G$ be a connected and undirected graph. Let $n$ denote the number of vertices of $G$.
We now show that for any pair of nodes $i,j$, such that $\mathrm{degree}(i)>\mathrm{degree}(j)$ there exists a $\delta^{ij} > 0$ such that $C^H_i(t) > C^H_i(t)$ whenever $0 < t <\delta^{ij}$. Since we have a finite number of pairs $i,j$ we can then take $\delta$ to be the smallest among the $\delta^{ij}$, making the result valid for all pairs $i,j$ satisfying the hypothesis.

Given two nodes $i,j$ such that $\mathrm{degree}(i)>\mathrm{degree}(j)$, let $(C^H_i(t))'$ denote que derivative of $C^H_i(t)$ with respect to $t$. We first show that there exists $\delta^{ij} > 0$ such that $(C^H_i(t))'>(C^H_j(t))'$ whenever $0<t<\delta^{ij}$, by showing that
\be
\lim_{t  \to 0} \frac{(C^H_i(t))'}{(C^H_j(t))'} = \frac{\mathrm{degree}(i)}{\mathrm{degree}(j)} > 1\,.
\ee
It follows from the definition of a limit that there must exist a $\delta^{ij} > 0$ such that
\be
\frac{(C^H_i(t))'}{(C^H_j(t))'} > 1, \forall ~~0 < t < \delta^{ij}.
\ee 
It can be easily seen that
\be
\label{der_ch}
-(C^H_i(t))' = \sum_{l=1}^n \log_2(p_l(t|i))p'_l(t|i).
\ee
This follows by direct application of the chain rule and using the fact that $\sum_{l=1}^N p'_l(t|i) = 0$.

Let $L^{(k)}$ denote the kth power of $L$ the laplacian matrix of $G$. Using the series expansion of $p(t|i) = e^{-tL}p(0|i)$ and letting $I$ denote the identity matrix, we see that
\be
p_l(t|i) = I_{li} - t L_{li} + \frac{t^2}{2!}L^{(2)}_{li} + {\cal{O}}(t^3).
\ee
Thus,
\be
\label{exp_p}
p'_l(t|i) = -L_{li} + tL^{(2)}_{li} + {\cal{O}}(t^2).
\ee
Let $N(i)$ be the set of vertices of $G$ which are directly connected to $i$.
 The sum in $\eqref{der_ch}$ can be conveniently rewritten in three parts:

\begin{multline}
\label{3_terms}
-(C^H_i(t))'  =  \log_2(p_i(t|i))p'_i(t|i) + \\   +\sum_{\substack{l \in N(i)}} \log_2(p_l(t|i))p'_l(t|i) +\\
+\sum_{\substack{l \not\in N(i),  l \neq i }} \log_2(p_l(t|i))p'_l(t|i). 
\end{multline}
Using the expansion in \eqref{exp_p}, we have that if $l \in N(i)$ and $t$ is sufficiently small, 
\be
\begin{split}
    \log_2(p_l(t|i))p'_l(t|i) &= \log_2(t + {\cal{O}}(t^2))(1 + {\cal{O}}(t)) \\ &\approx \log_2(t).
\end{split}
\ee
Thus, as $t \to 0$, the middle term in \eqref{3_terms} grows as
\be\label{mid_term}
\begin{split}
\sum_{\substack{l \in N(i)}} \log_2(p_l(t|i))p'_l(t|i) &\approx \sum_{\substack{l \in N(i)}} \log_2(t) \\ &=\mathrm{degree}(i)\log_2(t).
\end{split}
\ee
Also due to the expansions, it can be shown that the first and last terms in \eqref{3_terms} go to $0$ as $t \to 0$. This together with \eqref{mid_term} imply that,
\be
\label{eq:sandrafalou}
\lim_{t \to 0} \frac{-(C^H_i(t))'}{\log_2(t)} = \mathrm{degree}(i).
\ee
The same applies to $(C^H_j(t))'$, from which we can conclude that as $t \to 0$,
\be 
\lim_{t \to 0} \frac{(C^H_i(t))'}{(C^H_j(t))'} =  \frac{\mathrm{degree}(i)}{\mathrm{degree}(j)} > 1,
\ee
and so there exists $\delta^{ij} > 0$ such that whenever $0<t<\delta_{ij}$, $(C^H_i(t))' > (C^H_j(t))'$. This implies that
\be
\begin{split}
C^H_i(t) = \int_{0}^t (C^H_i(t))' dt > \int_{0}^t (C^H_j(t))' = C^H_i(t),
\end{split}
\ee
whenever $0<t<\delta_{ij}$. This completes the proof.\hfill$\square$

To obtain an approximate formula for $C^H_i(t)$ for  $t\rightarrow 0$ one may proceed by integrating the approximation given in equation (\ref{eq:sandrafalou}) for $(C^H_i(t))'$ between $0$ and $t$ :
\be
C^H_i(t) \approx t ~ \mathrm{degree}(i) \left(\log_2(t) - \frac{1}{\ln(2)}\right)
\ee
This result can be used as a benchmark for numerical computations.


\vskip 0.5cm
\begin{thebibliography}{10}
%
%
\bibitem{BORGATTI200555}
S. P. Borgatti, Social Networks {\bf 27}, 55 (2005).
%
%
\bibitem{ilprints422}
L. Page, S. Brin, R. Motwani, and T. Winograd,Technical  Report  1999-66, SIDL-WP-1999-0120 (Stanford InfoLab, Nov. 1999)
%
%
%
\bibitem{Delvenne_2011}
J.-C. Delvenne and A.-S. Libert, Phys. Rev. E {\bf 83},046117 (2011).
%
%
\bibitem{TUTZAUER2007249}
F. Tutzauer, Social Networks {\bf 29}, Special Section: Advances in Exponential Random Graph (p*) Models, 249 (2007).
%
%
\bibitem{NIKOLAEV2015154} 
A. G. Nikolaev, R. Razib, and A. Kucheriya, Social Networks {\bf 40}, 154 (2015).
%
%
\bibitem{split&transfer}
F. Oggier, S. Phetsouvanh, and A. Datta, PeerJ Computer Science {\bf 5}, e220 (2019).
%  
% 
\bibitem{PhysRevResearch.2.033104}
A.  Arnaudon,  R.  L.  Peach,  and  M.  Barahona,  Phys. Rev. Research {\bf 2}, 033104 (2020).
%
%
\bibitem{DiffusionMaps} 
R. R. Coifman, S. Lafon, A. B. Lee, M. Maggioni, B.Nadler,  F.  Warner,  and  S.  W.  Zucker,  Proceedings  of the National Academy of Sciences {\bf 102}, 7426 (2005).
%
%
\bibitem{Stability_Com} 
 J.-C. Delvenne, S. N. Yaliraki, and M. Barahona, Proceedings  of  the  National  Academy  of  Sciences {\bf 107},12755 (2010).
%
%
\bibitem{zoom_lens}. 
  M. T. Schaub, J.-C. Delvenne, S. N. Yaliraki, and M.Barahona, PLOS ONE {\bf 7}, 1 (2012).
%
%
\bibitem{THOMASCOVER}
T. M. Cover and J. A. Thomas, {\it Elements of information Theory}, 2nd ed. (John Wiley  Sons, Ltd, 2006).
%
%
\bibitem{Katz1953}
L. Katz, Psychometrika {\bf 18}, 39 (1953).
%
%
\bibitem{matrix_functions}
E.  Estrada  and  D.  J.  Higham,  SIAM  Review {\bf 52},  696 (2010).
%
%
\bibitem{subgraph_cent} 
E. Estrada and J. A. Rodriguez-Velazquez, Phys. Rev.E {\bf 71}, 056103 (2005).
%
%
\bibitem{total_comunicability}
M. Benzi and C. Klymko, Journal of Complex Networks {\bf 1}, 124 (2013).
%
%
\bibitem{Boccaletti2006}
S.  Boccaletti et  al.,  Physics  Reports {\bf 424},  175(2006).
%
%
\bibitem{Newman2003} 
M. Newman, SIAM Review {\bf 45}, 167 (2003).
%
%
\bibitem{Newman2001}.
M. Newman, Physical Review E {\bf 64}, 016132 (2001).
%
%
\bibitem{matrix_limits}
T. Valente, K. Coronges, C. Lakon, and E. Costenbader, Connections (Toronto, Ont.) {\bf 28}, 16 (2008).
%
%
\bibitem{correl}
M.  Benzi  and  C.  Klymko,  SIAM  Journal  on  Matrix Analysis and Applications {\bf 36}, 686 (2015).
%
%
\bibitem{alpha_cent}
P.  Bonacich  and  P.  Lloyd,  Social  Networks {\bf 23},  191(2001)
  
\end{thebibliography}
\end{document}